\documentclass[aps,reprint,prl]{revtex4-1}

\usepackage{bm}
\usepackage{graphicx} %
\usepackage{amsmath} %
\usepackage{amssymb} %
\usepackage{url}
\usepackage{subfigure} %
\usepackage{float} %
\usepackage{upgreek} %
\usepackage{multirow} %
\usepackage{color} %
\usepackage{lineno} %
\usepackage{hyperref} %
\usepackage{booktabs}
\usepackage{placeins}
\usepackage{autobreak}
\usepackage[usenames,dvipsnames]{xcolor}
\hypersetup{colorlinks=true, urlcolor=black, linkcolor=blue, citecolor=red} %
\allowdisplaybreaks[4]

\usepackage{ulem}

\begin{document}

\title{Statistical anisotropy in CMB spectral distortions}

\author{Atsuhisa Ota}
\email{a.ota@damtp.cam.ac.uk}
\affiliation {Department of Applied Mathematics and Theoretical Physics, University of Cambridge, Cambridge, CB3 0WA, UK}
\affiliation{Institute for Theoretical Physics and Center for Extreme Matter and Emergent Phenomena,
Utrecht University, Princetonplein 5, 3584 CC Utrecht, The Netherlands}
\date{\today}

\begin{abstract}
Measurements of the cosmic microwave background (CMB) spectral $y$-distortion anisotropy offer a test for the statistical isotropy of the primordial density perturbations on $0.01\lesssim k{\rm Mpc}\lesssim 1$.  
We compute the \textit{1-point ensemble averages} of the $y$-distortion %\sout{harmonic coefficients $\langle y_{\ell m}\rangle$}\AO{
anisotropies %}
 which vanish for the statistically isotropic perturbations. %\sout{ except for $\ell=0$.}
For the quadrupole %\AO{
statistical %}
anisotropy, we find $4\pi\langle y_{2m}\rangle=-6.8A_2\times 10^{-9}Y_{2m}(\mathbf d )$ with the quadruple Legendre coefficient of the anisotropic powerspectrum $A_2$ and the $\ell=2$ spherical harmonics $Y_{2m}(\mathbf d )$ for the preferred direction $\mathbf d $.
Also, we %\AO{
discuss %}\sout{argue}
the cosmic variance of the  $y$-distortion anisotropy in the statistically anisotropic Universe.
%\pacs{04.80.Cc, 95.30.Sf, 98.70.Vc, 98.80.Es}
\end{abstract}

\maketitle

%% MAIN TEXT %%%%%%%%%%%%%%%%%%%%%%%%%%%%%%%%%%%%%%%%%%%%%%%

There exist $10^{-5}$ of anisotropies in the cosmic microwave background~(CMB), the \textit{fossil} of the radiation emitted about 380,000 years after the Big Bang~\cite{Smoot:1992td,Hinshaw:2012aka,Aghanim:2018eyx}.
These fluctuations are random fields on top of the statistically isotropic background spacetime, and \textit{cosmic inflation} can explain their origin as the quantum fluctuations in the very early stage of the expanding Universe~\cite{Mukhanov:1981xt,Guth:1982ec,Hawking:1982cz}.
However, such a rotational invariance is not a mandatory requirement.
Indeed several inflationary models can break the rotational symmetry in the early Universe~\cite{Ackerman:2007nb,Soda:2012zm,Maleknejad:2012fw,Dimastrogiovanni:2010sm}.
For example, a vector field during inflation leads to a preferred direction and produces the quadrupole asymmetry in the primordial powerspectrum of the density perturbations~\cite{Soda:2012zm,Maleknejad:2012fw,Dimastrogiovanni:2010sm}.
More generally, spinning particles imprint the multipole asymmetry of the primordial correlators~\cite{Arkani-Hamed:2015bza,Bartolo:2017sbu,Franciolini:2017ktv}, and hence %\AO{
the statistical asymmetries are %}\sout{it is} 
sensitive to the matter contents during inflation.
Thus, the statistical isotropy is an assumption to be tested through observations, 
and its probes have been discussed with the powerspectra of the CMB anisotropies, the 21-cm lines and galaxies~\cite{Kim:2013gka,Ramazanov:2013wea,Tansella:2018hdm,Bartolo:2014hwa,Naruko:2014bxa,Ade:2015hxq,Shiraishi:2016omb,Ramazanov:2016gjl,Shiraishi:2016wec,Sugiyama:2017ggb,Durakovic:2017prf}.

In this \textit{Letter}, we point out another way to link the primordial statistical anisotropy with a cosmological observable, i.e., a deviation of the CMB energy spectrum from the blackbody one.
In particular, we discuss the sensitivity for the short wavelength~($\mathcal O(0.01)<k{\rm Mpc}<\mathcal O(1.)$) statistical anisotropy.
We focus on the spectral $y$-distortion, a kinetic deviation from the Planck distribution due to the Compton scattering in the late epoch of the early Universe~\cite{Zeldovich:1969ff,Sunyaev:1970er}.
A typical source of the $y$-distortion is energy release from dissipation of acoustic waves on small scales.
It produces the $y$-distortions at second-order in the cosmological perturbations, and hence the ensemble average of them is related to the short wavelength primordial powerspectrum~\cite{Hu:1994bz,Chluba:2012we,Chluba:2012gq}.
In particular, the isotropic spectral distortion has been studied for the small-scale primordial powerspectrum.
Here, we investigate imprints of the primordial statistical anisotropy on the $y$-distortion \textit{anisotropy}.
In contrast to Refs~\cite{Shiraishi:2015lma,Dimastrogiovanni:2016aul,Shiraishi:2016hjd}, we do not discuss the 2- or higher correlation functions of the spectral distortion anisotropy.
Instead, we compute a simple \textit{1-point ensemble average} of the $y$-distortion anisotropy in an explicit way based on the second-order Boltzmann equation.
Such a quantity is zero in the statistically isotropic Universe, but we show that it is sensitive to the statistical anisotropy.
Also, we comment on the cosmic variance in the statistically anisotropic Universe.

\medskip
\textit{Primordial powerspectrum---.}
Let $\mathbf d $ be the preferred direction in the Universe.
The powerspectrum of the curvature perturbation on the uniform density slice $\zeta$ depends on $\mathbf d $, and we write it in Fourier space as
\begin{align}
    \langle \zeta_{\mathbf k_1}\zeta_{\mathbf k_2} \rangle 
=(2\pi)^3 \delta^{(3)}(\mathbf k_1+\mathbf k_2)P_\zeta^{\mathbf d }(\mathbf k_1).\label{power}
\end{align}
The SO(3) rotational symmetry is broken to SO(2) in the presence of $\mathbf d $.
The Universe is still statistically symmetric under the rotation around the $\mathbf d $ axis; therefore, the anisotropy can be parameterized by the angle between $\mathbf d $ and $\mathbf k$.
Then, we generally expand the powerspectrum by the Legendre polynomial $P_{L}(\mathbf d \cdot \hat k)$ as 
\begin{align}
        P_\zeta^{\mathbf d }(\mathbf k) 
        =P_\zeta(k) 
        \sum_{L=0}^\infty 
        (-i)^L(2L+1) A_L(k) 
        P_L(\mathbf d \cdot {\hat k}),
        \label{def:power}      
\end{align}
where $\hat k \equiv \mathbf k/|\mathbf k|$, $k\equiv |\mathbf k|$, $P_\zeta$ is the isotropic part of the powerspectrum. 
In the statistically isotropic Universe, $A_0=1$ and $A_{L\neq 0}=0$.
 %\sout{Eq.~(\ref{power}) should be defined as symmetric under $1\leftrightarrow 2$ replacement.}
%\AO{
Also, $P_{\zeta}(\mathbf k)$ should be defined as symmetric under $\mathbf k\leftrightarrow -\mathbf k$.
%}
Then, the odd terms in Eq.~(\ref{def:power}) are zero.
Therefore, the simplest nontrivial asymmetry of the powerspectrum is $L=2$, which is motivated in anisotropic inflation~(For example, see Refs.~\cite{Ackerman:2007nb,Soda:2012zm,Maleknejad:2012fw,Dimastrogiovanni:2010sm}).
We assume $A_L$ has no scale dependence for simplicity.

\medskip 
\textit{Spectral distortions---.}
\label{multipoletrans}
Let $\eta$ be the conformal time.
We write the linear photon temperature perturbation at $\mathbf x$ as $\Theta(\eta, \mathbf x,\mathbf n)$.
Here, the unit vector $\mathbf n$ is the photons' direction. 
The harmonic coefficients of the temperature perturbation are defined as
\begin{align}
    \Theta_{\ell m}(\eta, \mathbf x)\equiv \int d\mathbf n Y^*_{\ell m}(\mathbf n)\Theta (\eta, \mathbf x, \mathbf n).
    \label{defHC}
\end{align}
The temperature perturbations are linear in $\zeta$ in Fourier space:
\begin{align}
    \Theta(\eta, \mathbf x,\mathbf n)
    =&\int \frac{d^{3}k}{(2\pi)^{3}}
    e^{i\mathbf k\cdot\mathbf x}
    \sum_{\ell m}(-i)^{\ell}    (4\pi)\notag \\
    &\times Y^{*}_{\ell m}(\mathbf n)
    Y_{\ell m}(\hat k)
    \Theta_{\ell}(\eta,k)
    \zeta_{\mathbf k}.
\end{align}
Let $\mathbf v(\eta, \mathbf x)$ is the velocity of the baryon fluid at $\mathbf x$.
Then we define $V(\eta, \mathbf x,\mathbf n)\equiv\mathbf n\cdot \mathbf v(\eta, \mathbf x)$.
Note that we reserved capital letters $L$ and $M$ for the primordial statistical asymmetry.

The spectral $y$-distortion is a deviation from the blackbody spectrum at second-order in the cosmological perturbations.
From the photon Boltzmann equation at second-order, the $y$-distortion obeys~\cite{Pitrou:2009bc,Chluba:2012gq,Chluba:2016aln,Ota:2016esq}
\begin{align}
    &\dot y +  \mathbf n\cdot  \nabla  y =-\dot \tau S^{(\rm ac.)} -\dot \tau \frac{T_{\rm e}-T_0(1+z)}{m_{\rm e}}\notag \\
    +&\dot \tau \left(y-\frac{1}{\sqrt{4\pi}}y_{00}- \frac{1}{10}\sum_{m=-2}^2Y_{2m}y_{2m}\right),
    \label{y:eom:real}
\end{align}
where $\tau$ is the optical depth~($\dot\tau<0$), $T_{\rm e}$ is the electron temperature, $m_{\rm e}$ is electron mass, $T_0=2.725$K, $z$ is the redshift, and the over-dots are the partial derivative w.r.t. the conformal time.
$y_{\ell m}$ is defined in the same way with Eq.~(\ref{defHC}).
Here we introduced the second-order acoustic source~\cite{Pitrou:2009bc,Chluba:2012gq,Ota:2016esq}
\begin{align}
S^{(\rm ac.)} =
&\frac{1}{2}(V-\Theta)^2
+\frac{1}{\sqrt{4\pi}}\Theta_{00}(V-\Theta)\notag \\
&
+\frac{1}{10}(V-\Theta)\sum_{m=-2}^{2}Y_{2m}\Theta_{2m}
\notag \\
&
+\frac{1}{2\cdot \sqrt{4\pi}}[(V-\Theta)^2]_{00}\notag \\
&
+\frac{1}{20}\sum_{m=-2}^{2}Y_{2m}[(V-\Theta)^2]_{2m}
.
\label{source_y}
\end{align}
Eq.~(\ref{y:eom:real}) is valid for the low redshift~($z<5\times 10^4$) where the energy transfer due to the baryons is negligible~\cite{Hu:1992dc,Chluba:2013vsa}.
We ignore such high energy corrections to Eq.~(\ref{y:eom:real}) since we are interested in the period of recombination~($z\sim 10^3$) and reionization~($z\sim 10$) in the following discussions.   

The homogeneous and isotropic component of Eq.~(\ref{y:eom:real}) has the form
\begin{align}
    \dot y_{00}= -\dot\tau S^{(\rm ac.)}_{00}-\sqrt{4\pi}\dot \tau \frac{T_{\rm e}-T_0(1+z)}{m_{\rm e}}.
\end{align}
This is nothing but the evolution equation for the spectral $y$-distortion derived in the previous literature.
Taking the ensemble average of both sides and integrating with respect to time, we obtain~\cite{Chluba:2012gq}
\begin{align}
    \langle y_{00} \rangle =&\sqrt{4\pi}  \int_{\eta_i}^{\eta_0}d\eta (-\dot \tau)
 \int \frac{dk}{k}\frac{k^{3}}{2\pi^{2}}P_{\zeta}
\left[
3\Theta_{1g}^2
+
\frac{9}{2}\Theta_2^2
+\cdots
 \right]\notag \\
 &+\sqrt{4\pi}\int_{\eta_i}^{\eta_0}d\eta (-\dot \tau)
 \frac{T_{\rm e}-T_0(1+z)}{m_{\rm e}}
 ,
 \label{isoydis}
\end{align}
where the first line is the acoustic energy injection and the second one is the Sunyaev-Zel'dovich effect.
%\AO{
We have also introduced relative velocity $\Theta_{1g}=\Theta_1-V_1$.
%}
In this \textit{Letter}, we are more interested in the homogeneous and \textit{anisotropic} components of the $y$-distortion.
Dropping the gradient term and taking the $\ell=2$ component of the ensemble average of Eq.~(\ref{y:eom:real}), we find
\begin{align}
        \left \langle \dot y_{2m}\right\rangle =\dot \tau \frac{9}{10}\left \langle y_{2m}\right \rangle -\dot \tau \left \langle S^{(\rm ac.)}_{2m}\right\rangle .\label{eom:bolt:y}
\end{align}
The first term in Eq.~(\ref{eom:bolt:y}) implies that the Thomson scattering exponentially suppresses $\langle y_{2m}\rangle $ without sources.
To calculate the harmonic coefficients of the second-order acoustic source, we use the following formula:
\begin{align}
    &  \left \langle \int d\mathbf n Y^*_{\ell m}(\mathbf n) \prod_{i=1}^{2}B^{(i)}(\mathbf x,\mathbf n,\mathbf d )\right \rangle\notag \\ 
=&    
\sum_{Ll_{1}l_{2}Mm_{1}m_{2}}(-i)^{l_{1}-l_{2}+L}
Y^*_{LM}(\mathbf d )\mathcal G^{*\ell,l_1,l_2}_{m,m_1,m_2}
\mathcal G^{L,l_1,l_2}_{M,m_1,m_2}
\notag\\
&
\times (4\pi)^2\int \frac{dk }{k }\frac{k ^{3}}{2\pi^{2}}A_L P_{\zeta}(k ) 
B^{(1)}_{l_1}(k )B^{(2)}_{l_2}(k ),
\label{kousiki}
\end{align}
where the linear perturbations $B^{(i)}$~($i=1,2$) are expanded into
\begin{align}
    B^{(i)}(\mathbf x,\mathbf n,\mathbf d )=&\int \frac{d^{3}k_{i}}{(2\pi)^{3}}e^{i\mathbf k_{i}\cdot\mathbf x}\sum_{l_{i}m_{i}}(-i)^{l_{i}}(4\pi)\notag \\
    &\times Y^{*}_{l_{i}m_{i}}(\mathbf n)Y_{l_{i}m_{i}}(\hat k_{i})B^{(i)}_{l_i}(k_{i})\zeta_{\mathbf k_{i}}.
\end{align}
We also introduced the Gaunt integral $\mathcal G^{\ell_1,\ell_2,\ell_3}_{m_1,m_2,m_3}\equiv \int  d\mathbf n Y_{\ell_1 m_1}(\mathbf n)Y_{\ell_2 m_2}(\mathbf n)Y_{\ell_3 m_3}(\mathbf n).$
Using Eqs.~(\ref{kousiki}) to (\ref{source_y}) with (\ref{def:power}) and taking $\ell=2$, we find
\begin{align}
    \left \langle 
S^{(\rm ac.)}_{2m} \right \rangle=&
4\pi Y^*_{2m}(\mathbf d ) \int \frac{dk }{k }\frac{k ^{3}}{2\pi^{2}}(-i)^{2}A_2 P_{\zeta}
\notag \\
&\times \left[
\frac{33}{50}\Theta_{1g}^2
+
\frac{9}{14}\Theta_2^2
+\cdots
 \right].
 \label{acs}
\end{align}
%\sout{where we have introduced relative velocity $\Theta_{1g}=\Theta_1-V_1$.}
The dots imply the higher order multipoles, which we ignore on the analogy of the isotropic acoustic source~\cite{Chluba:2012gq}.
Thus, there exists a nonzero anisotropic acoustic source in case of the statistically anisotropic Universe.
Eq.~(\ref{acs}) is composed of heat conduction $\Theta_{1g}$ and shear viscosity $\Theta_2$ of the photon-baryon plasma; therefore, the quadrupole of the acoustic source is gauge invariant.
Note that $\langle y_{2m}\rangle$ is initially zero, and hence it remains zero if $A_2=0$.
It is possible to formally integrate Eq.~(\ref{eom:bolt:y}):
\begin{align}
    \left \langle y_{2m}\right\rangle =\int^\eta_{\eta_i}d\bar \eta 
    \left[ -\dot \tau e^{-\frac{9}{10}[\tau-\tau(\eta)]}
\right]    \left \langle S^{(\rm ac.)}_{2m}\right\rangle,
\end{align}
where the arguments are $\bar\eta$ unless otherwise stated. 
Combining this expression with Eq.~(\ref{acs}), we write the anisotropy at present $\eta=\eta_0$ as follows: 
\begin{align}
    \left \langle y_{2m}\right\rangle\Big|_{\eta=\eta_0} =&4\pi Y^*_{2m}(\mathbf d )\int^{\eta_0}_{\eta_i}d\eta 
    g_{\frac{9}{10}}
 \int \frac{dk }{k }\frac{k ^{3}}{2\pi^{2}}
 \notag \\
    &\times (-i)^{2}A_2 P_{\zeta}
\left[
\frac{33}{50}\Theta_{1g}^2
+
\frac{9}{14}\Theta_2^2
+\cdots
 \right]
    \label{y2mres},
\end{align}
where we have used $\tau(\eta_0)=0$ and defined the visibility function which picks the contribution around the last scattering up:
\begin{align}
    g_{\frac{9}{10}}(\eta) \equiv -\dot \tau(\eta) e^{-\frac{9}{10}\tau(\eta)}.
\end{align}
The anisotropic acoustic source has the visibility function in contrast to Eq.~(\ref{isoydis}).
This implies that $\langle y_{2m}\rangle$ is generated during recombination and reionization only.
\begin{figure}[t]
  \includegraphics[width=80mm]{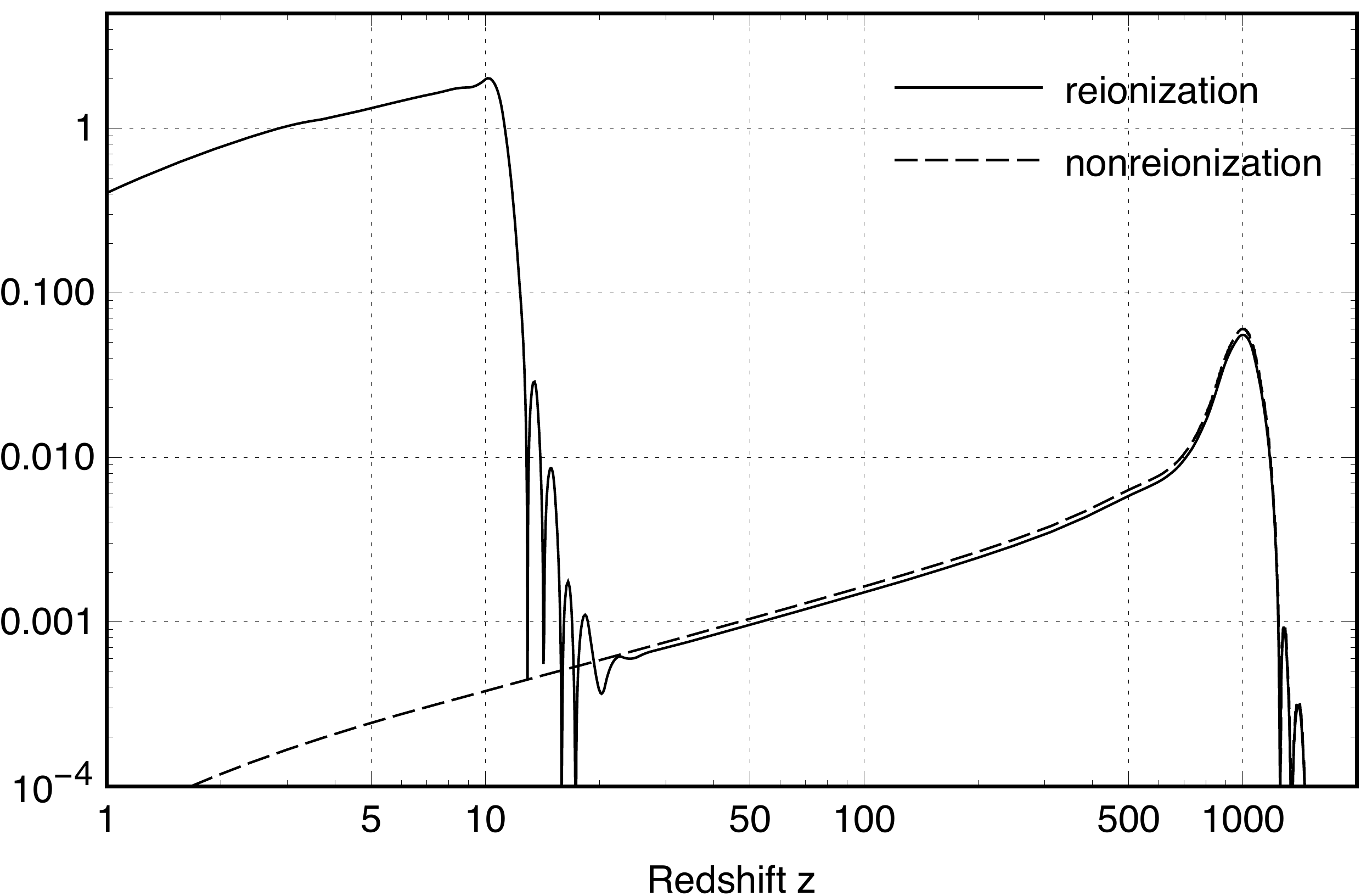}
  \caption{The $\ell=2$ heating rate $d\langle y_{2m}\rangle/d\ln z$ in units of $4\pi A_2 A_\zeta Y_{2m}$.}
  \label{win_z}
\end{figure}
\begin{figure}[t]
  \includegraphics[width=80mm]{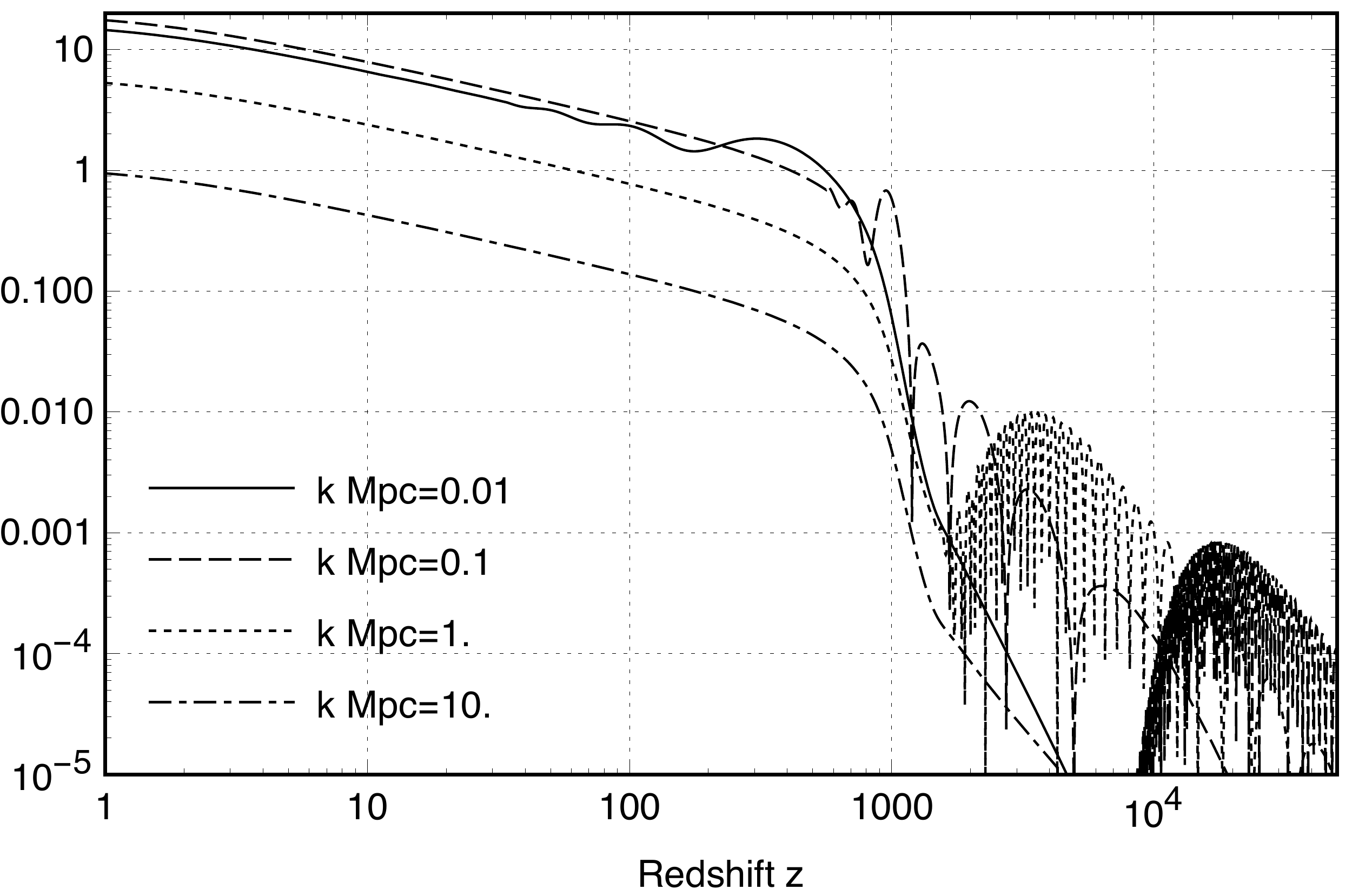}
  \caption{Time evolution of $3\Theta_{1g}$ of various Fourier momenta.}
  \label{bv}
\end{figure}
In Fig.~\ref{win_z}, we show the numerical estimation of $d\langle y_{2m}\rangle/d\ln z$ by using the Cosmic Linear Anisotropy Solving System~(\texttt{CLASS})~\cite{Blas:2011rf}.
We model the isotropic component of the powerspectrum as $k^{3}(2\pi^{2})^{-1}P_\zeta(k)=A_\zeta (k/k_0)^{n_s-1}$ with $10^{9}A_\zeta=2.196$, $n_s=0.96$ and $k_0{\rm Mpc}=0.05$.
The figure shows that $g_{\frac{9}{10}}$ suppresses the contribution from $z\gg 10^3$.
Also, $\langle y_{2m} \rangle$ is generated not only during recombination but also during reionization, and the dominant contribution comes from the latter.
This is because the baryon bulk velocity significantly grows after recombination so that $\Theta_{1g}$ is enhanced as we show in Fig.~\ref{bv}.
Indeed, we found the bulk motion of the ionized baryons after the reionization enhances the homogeneous and isotropic component of the $y$-distortion about ten times~\cite{ota:prep}.
%\sout{We directly see the quadrupole anisotropy on the last scattering surface as depicted in the left panel in Fig.~\ref{fig1}, but the anisotropy before the last scattering is erased by the Thomson scattering as the right figure shows.}
%\AO{
The role of the visibility function in Eq.~(\ref{y2mres}) is explained as follows: we directly see the quadrupole anisotropy on the last scattering surface as depicted in the left panel in Fig. 3, 
but the anisotropy before the last scattering epoch is erased by the Thomson scattering as the right figure shows.
%}

We also see the $L=4$ primordial anisotropy in the $\ell=4$ anisotropy.
The similar calculation yields the $\ell=4$ $y$-distortion of the form
\begin{align}
    \left \langle 
y_{4m} \right \rangle\Big|_{\eta=\eta_0}=&
4\pi Y^*_{4m}(\mathbf d ) \int^{\eta_0}_{\eta_i}d\eta 
    g \notag \\
    &\times \int \frac{dk }{k }\frac{k ^{3}}{2\pi^{2}}A_4 P_{\zeta}
 \left[
\frac{5}{7} \Theta_2^2+\cdots
 \right],
 \label{l4d}
\end{align}
where we defined $g=-\dot \tau  e^{-\tau}$.
Extension to the higher $L=\ell$ is possible if we account for the higher multipole moments such as $\Theta_{3}$ which we ignore here.
\begin{figure}[t]
\includegraphics[width=70mm]{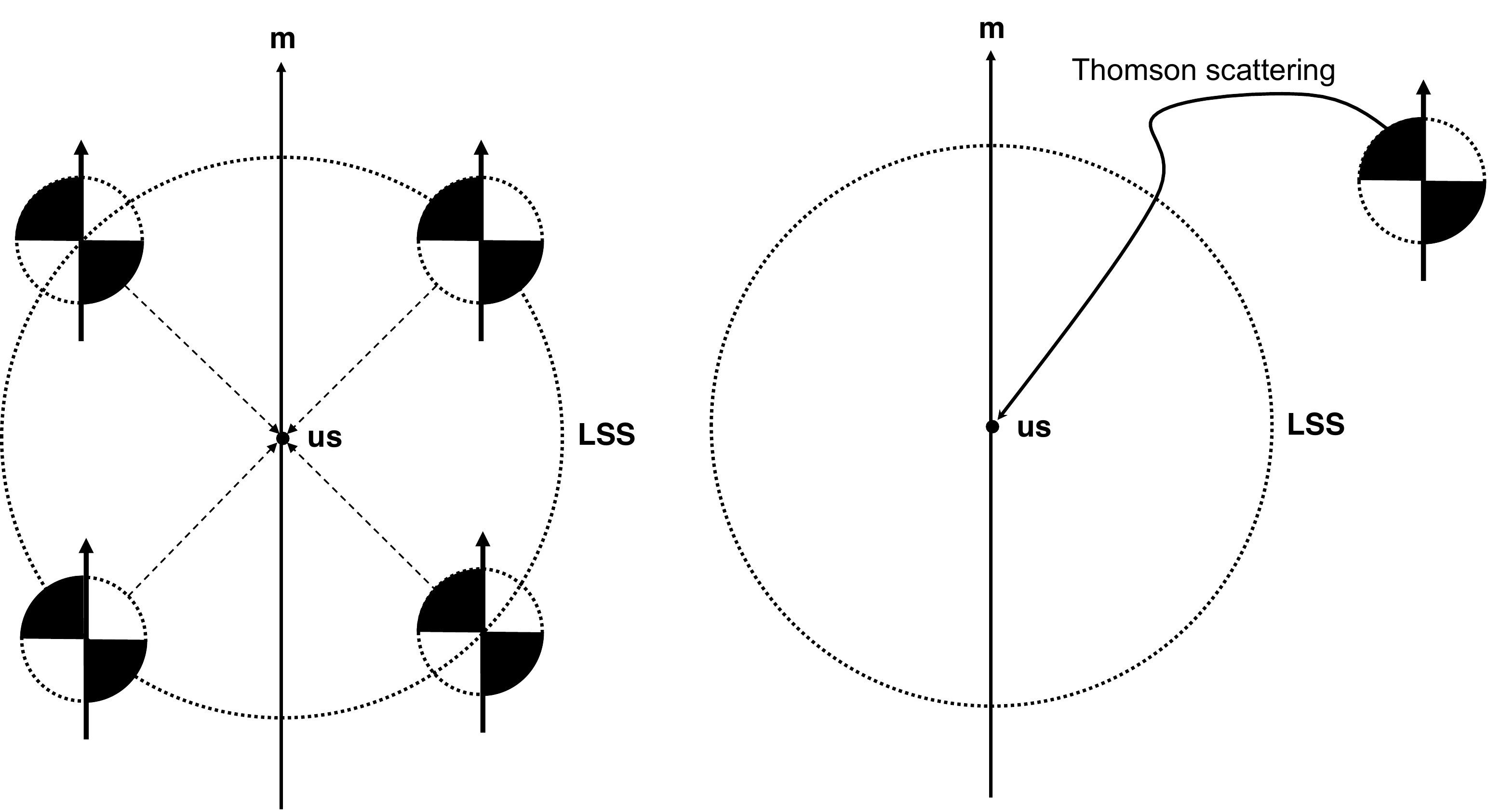}
  \caption{The quadrupole asymmetry on the LSS.}
  \label{fig1}
  \end{figure}

\begin{figure}[t]
  \includegraphics[width=80mm]{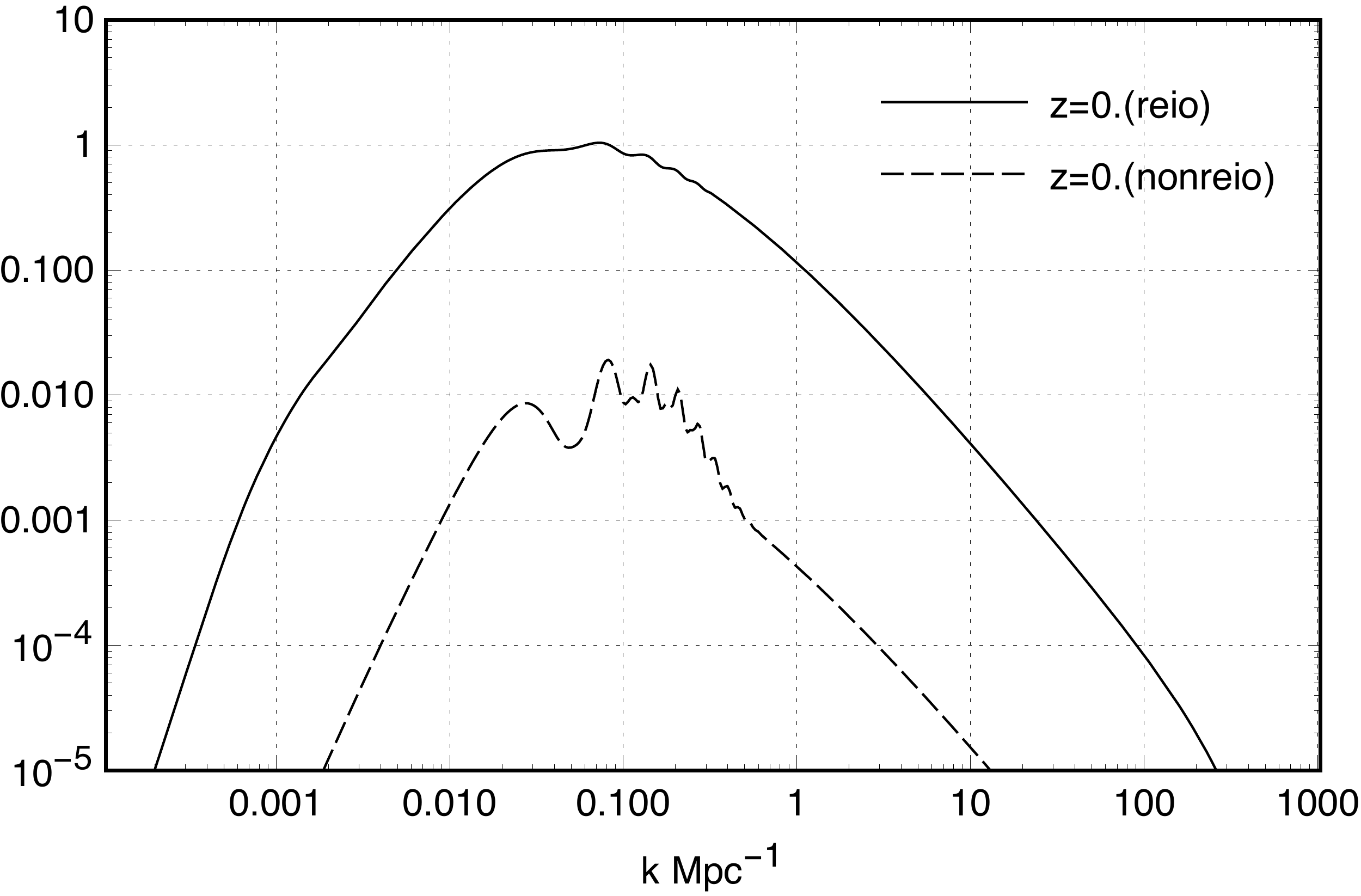}
  \caption{$d \alpha_2/d\ln k $ in Fourier space in units of $|A_2|A_\zeta$.
  }
  \label{win_K}
\end{figure}

\medskip
\textit{Observables---.}
The observed $y$-distortion anisotropy is expanded by $Y_{\ell m}$ and is encoded into $(2\ell +1)$ values.
Eq.~(\ref{y2mres}) contains the unknown vector $\mathbf d $ so that each of $y_{2 m}$ depends on the choice of the observer's axis.
We accordingly have to introduce the coordinate-independent quantity
\begin{align}
    \alpha_\ell \equiv \frac{1}{4\pi}\sqrt{\frac    {4\pi}{2\ell +1}\sum_{m=-\ell}^{\ell}|\langle y_{\ell m}\rangle |^2},
    \label{def:alpha}
\end{align}
where the normalization factor is chosen to satisfy $\alpha_0=\sqrt{4\pi}^{-1} \langle y_{00}\rangle$.
For $\ell=2$, combining Eqs.~(\ref{y2mres}) with (\ref{def:alpha}), we obtain 
\begin{align}
    \alpha_{2} = \int_{\eta_i}^{\eta_0}d\eta 
    g_{\frac{9}{10}}    
 \int \frac{dk }{k }\frac{k ^{3}}{2\pi^{2}}|A_2| P_{\zeta}
\left[
\frac{33}{50}\Theta_{1g}^2
+
\frac{9}{14}\Theta_2^2
+\cdots
 \right]
    \label{alpha2mres}.
\end{align}
We numerically estimated $\alpha_2$ with reionization and find
\begin{align}
    \alpha_2 =6.8|A_2|\times 10^{-9}.
\end{align}
If we ignore the contribution from reionization, we obtain 
\begin{align}
         \alpha^{\rm NR}_2  =6.0|A_2|\times 10^{-11}.
\end{align}
Thus, the reionization is dominant as we have already seen in Fig.~\ref{win_z}.
$\alpha_2$ is the integrated quantity of the powerspectrum on some scales.
In Fig.~\ref{win_K}, we illustrate the $k$ logarithmic derivative of $\alpha_2$ at $z=0$.
We found $\alpha_2$ is sensitive to the $A_2$ on scales $0.01\lesssim k{\rm Mpc}\lesssim 1.$, which corresponds to $10^2 \lesssim\ell \lesssim10^4$ in multipole of the temperature anisotropy.
Note that the reionization enhancement does not happen on the horizon scale during reionization.
This is because the enhancement comes from that of the baryon velocity $V$.
For $k {\rm Mpc} \gtrsim 0.1$, $V$ is exponentially suppressed due to Silk damping as illustrated in Fig.~\ref{bv}.
On the other hand, $k {\rm Mpc} \lesssim 0.1$ modes can grow after recombination due to the gravitational instability.
On larger scales, the velocity perturbations are not generated.
Similarly, we find the $L=4$ component from Eqs.~(\ref{l4d}) and (\ref{def:alpha}) as follows:
\begin{align}
    \alpha_{4} = \int_{\eta_i}^{\eta_0}d\eta 
    g\int \frac{dk }{k }\frac{k ^{3}}{2\pi^{2}}|A_4| P_{\zeta}
 \left[
\frac{5}{7} \Theta_2^2+\cdots
 \right].
\end{align}
We also estimated $\alpha_4$ as
\begin{align}
    \alpha_4=1.2|A_4|\times 10^{-11}.
\end{align}
For $L=4$, the enhancement of reionization is a percent level since it does not contain $\Theta_{1g}$.
Once we get $\alpha_{\ell}$, we can reconstruct the spherical harmonics as $Y_{\ell m}=(4\pi)^{-1}\langle y_{\ell m}\rangle\alpha^{-1}_\ell$ and can find the direction $\mathbf d $.

\medskip
\textit{Cosmic variance of the spectral distortions---.}
For $\ell=2$, we only have five samples so that one may wonder that our observable suffers from the sizeable cosmic variance on the analogy of the temperature anisotropy.
We give the theoretical prediction of the observables by taking the ensemble average, which corresponds to the statistical average of many quantum realizations.
In case of the statistically isotropic Universe, we can think of the different directions as different realizations of quantum fluctuations; therefore, we identify the observed angular average with the ensemble average:
\begin{align}
    \langle y^{\rm theory}(\mathbf n)\rangle =\int \frac{d\mathbf n}{4\pi}y^{\rm obs.}(\mathbf n).
\end{align}
We should calculate the RHS after discretizing the celestial sphere.
The typical patch size is given by the last diffusion scale of $y$-era $k^{-1}_{\rm D}=\mathcal O(10)$Mpc.
Hence, the number of samples is given by a fraction of the present horizon scale and $k^{-1}_{\rm D}$.
We roughly calculate this number as $(3000\cdot k_{\rm D})^2\sim 10^{5}$~\cite{Pajer:2012vz}.
Therefore, the cosmic variance of the spectral $y$ distortion is negligibly small.
For the statistically anisotropic case, the preferred direction decreases the number of samples because the various cosines from the preferred directions are no more equivalent.
In this case, we identify only the different azimuthal angle $\phi$ as the different quantum realization.
Then, the number of samples approximately becomes the square root of the number of diffusion patches on the sky, i.e., $3000\cdot k_{\rm D}\sim 300$.
Therefore, we observationally obtain $y_{2m}$ as the average of about 300 realizations, and the cosmic variance $\delta y_{2m}/y_{2m}$ is typically $\sim$10$\%$.

\medskip
\textit{Discussions---.}
We computed 1-point ensemble averages of the $y$-distortion anisotropies $\langle y_{\ell m}\rangle $ in cosmological perturbation theory.
Such quantities are zero for the statistically isotropic perturbations.
However, we found that the statistical anisotropies produce the nonzero contributions. 
One may wonder if we obtain the similar results for the spectral $\mu$-distortion, the chemical potential type spectral distortion generated during $5\times 10^4<z<2\times 10^6$.
However, the Compton scattering is efficient enough to erase the intrinsic angular dependence of $\mu$-distortion when it realizes the kinetic equilibrium; therefore, we cannot see any primordial statistical anisotropy in the $\mu$-distortion.
We used the linear Boltzmann solver to follow the evolution of $\Theta_{1g}$ and $\Theta_2$, while Fig.~\ref{win_K} suggests there is a contribution from the nonlinear scale.
We expect our results will be updated when we account for the nonlinear evolution of the matter perturbations.
The astrophysical background of the spectral $y$-distortion is more complicated compared to that from the primordial perturbations.
The Sunyaev-Zel'dovich effects from galaxy clusters may contaminate the signal from the primordial statistical anisotropy.
Hence, the masking techniques should be developed for data analysis.
The extension to more general statistical anisotropy is straightforward.
This should be investigated in the future works.

%% ACKNOWLEDGEMENT %%%%%%%%%%%%%%%%%%%%%%%%%%%%%%%%%%%%%%%%%%%%%%%
\begin{acknowledgments}
The author is supported by JSPS Overseas Research Fellowships.
We thank Enrico Pajer and Jens Chluba for useful discussions and comments.
The author is grateful to an anonymous referee of the Physics Letters B for careful reading and the useful comments.

\end{acknowledgments}

%\bibliography{bib}{}
%\bibliographystyle{unsrt}

\end{document}